\documentclass[reprint,twocolumn,amsmath,superscriptaddress,amssymb,aps,prb]{revtex4-1}

\usepackage{graphicx}% Include figure files
\usepackage{graphics}
\usepackage{amsmath}
\usepackage{dcolumn}% Align table columns on decimal point
\usepackage{bm}% bold math

\begin{document}

%\preprint{APS/123-QED}

\title{Biaxial-stress-driven full spin polarization in ferromagnetic hexagonal chromium telluride}

\author{Xiang-Bo Xiao}
\affiliation{Beijing National Laboratory for Condensed Matter Physics, Institute of Physics, Chinese Academy of Sciences, Beijing 100190, China}
\author{Jun Li}
\affiliation{Beijing National Laboratory for Condensed Matter Physics, Institute of Physics, Chinese Academy of Sciences, Beijing 100190, China}
\author{Bang-Gui Liu}\email{bgliu@iphy.ac.cn}
\affiliation{Beijing National Laboratory for Condensed Matter Physics, Institute of Physics, Chinese Academy of Sciences, Beijing 100190, China}
\affiliation{School of Physical Sciences, University of Chinese Academy of Sciences, Beijing 100190, China}

\date{\today}% It is always \today, today,

\begin{abstract}
It is important to spintronics to achieve fully-spin-polarized magnetic materials that are stable and can be easily fabricated. Here, through systematical density-functional-theory investigations, we achieve high and even full spin polarization for carriers in the ground-state phase of CrTe by applying tensile biaxial stress. The resulting strain is tensile in the xy plane and compressive in the z axis. With the in-plane tensile strain increasing, the ferromagnetic order is stable against antiferromagnetic fluctuations, and a half-metallic ferromagnetism is achieved at an in-plane strain of 4.8\%. With the spin-orbit coupling taken into account, the  spin polarization is equivalent to 97\% at the electronic phase transition point, and then becomes 100.0\% at the in-plane strain of 6.0\%. These make us believe that the full-spin-polarized ferromagnetism in this stable and easily-realizable hexagonal phase could be realized soon, and applied in spintronics.
\end{abstract}

\maketitle

%\tableofcontents

\section{Introduction}

High spin polarization, especially full spin polarization,  of charge carriers is highly desirable in magnetic materials for spintronics applications\cite{spintronics}. Half-metallic ferromagnetic materials\cite{hm} are promising for this purpose because its spin polarization at the Fermi level, by definition, can become 100\%. Rutile CrO$_2$\cite{cro2,cro2a}, Heusler compounds such as NiMnSb\cite{hm,heus1,heusler}, and double perovskite oxides\cite{dpo1,dpo2,dpo3} can make the best single-phase half-metallic materials. Nearly 100\% spin polarization can be realized near the Fermi level in good samples of them\cite{cro2b1,cro2b2}. Half-metallic ferromagnetism was also predicted in metastable zincblende and rocksalt CrTe phases\cite{crte-lbg,aaa} and afterwards the zincblende phase was fabricated in the form of ultrathin epitaxial films\cite{crte-cubic1,crte-cubic2}. On the other hand, it has been a great challenge to make high-quality samples of these half-metallic materials for perfect spin polarization\cite{heus1,cro2c,cro2d}. Therefore, it is vital to realize high or full spin polarization for carriers in stable materials that can be easily synthesized or fabricated. Fortunately, stable hexagonal CrTe compound is a good candidate for this purpose because it is the ground-state phase of CrTe and exhibits typical ferromagnetic order with high Curie temperature of 340K and a magnetic moment near $4\mu_B$\cite{crte1,crte2}.
These motivates us to seek high and even full spin polarization in this stable hexagonal material through experimentally realizable approaches.

Here, we apply a biaxial stress on the hexagonal CrTe and use full-potential density-functional-theory methods\cite{dft1,dft2} to optimize its strained crystal structures and then systematically investigate the electronic structures by using an improved exchange-correlation functional. We find that a biaxial tensile stress can make the hexagonal CrTe become half-metallic and thus realize a high spin polarization of 97\% at an in-plane strain of 4.8\% and then achieve full spin polarization at a strain of 6.0\%. The mechanism for such high spin polarization is also clarified. Because the hexagonal CrTe as the ground-state phase is stable and can be easily fabricated, these results can stimulate more exploration in this direction towards practical spintronics applications. More detailed result will be presented in the following.

\section{Computational details}

Our calculations are done by using full-potential augmented plane wave plus local orbitals methods within the density-functional theory implemented in the Vienna package WIEN2K \cite{wien2k}. We take the Perdew-Burke-Enzerhoff approach to the generalized-gradient approximation (GGA) for the exchange-correlation functional \cite{pbe}. To get more accurate description of the energy bands, including energy gaps, we use a modified Becke-Johnson (mBJ) exchange potential \cite{mbj}. This mBJ functional has been proved to substantially improve description of electronic structures of semiconductors and insulators\cite{mbj}, and even half-metallic materials\cite{hm-mbj}. The scalar relativistic approximation is used to treat the relativistic effects except for the spin-orbit coupling. We use 2000 k points in the Brillouin zone. $R_{mt}\times K_{max}$ is set to 7.5, and the expansion is done up to $l_{max}=10$ in the muffin tins. The muffin-tin radii of Cr and Te atoms are set to achieve high accuracy. When the integrated charge distance per formula unit between input charge density and output charge density is less than $0.0001|e|$, where $e$ is the electron charge, the self-consistent calculation will be considered to be converged \cite{wien2k}.

\section{Results and discussion}

\subsection{Ground-state properties}

The ground-state phase of CrTe is a metallic ferromagnetic phase in the hexagonal nickel-arsenide structure with the space group P6$_3$/mmc (\#194), and experimental lattice constants are $a_e=3.998$\AA, $c_e=6.254$\AA  \cite{crte1,crte2}. In this structure, Cr ans Te atoms occupy the (0,0,0) and (0.333,0.667,0.25) sites, respectively. At first, we have optimized ferromagnetic hexagonal CrTe with nickel-arsenide structure to determine the equilibrium lattice constants and the $c/a$ ratio in terms of the total energies. The equilibrium lattice constants are equivalent to $a_0=4.129$ \AA{} and $c_0=6.292$\AA, and the $c/a$ ratio is 1.524. It is useful to define $\alpha_1$ and $\alpha_2$ to characterize the two Cr-Te-Cr bond angles with the two Cr atoms being in the same xy plane and in the z axis, respectively. We obtain $\alpha_1=92.6^\circ$ and $\alpha_2=66.8^\circ$. The total density of states and partial density of states projected in the atomic spheres and interstitial region are presented in Fig. 1. Also presented in Fig. 1 are the partial density of states projected at the Cr e$_g$ and t$_{2g}$ orbitals. The crystal field splitting is small for the Cr d states. It is clear that the hexagonal CrTe is a usual ferromagnetic phase, but there is an energy gap a little away from the Fermi level in the minority-spin channel. This special electronic structure near the Fermi level implies that the hexagonal CrTe could be manipulated to enhance its spin polarization at the Fermi level.

\begin{figure}[!htbp]
\centering  % Requires \usepackage{graphicx}
\includegraphics[clip, width=7cm]{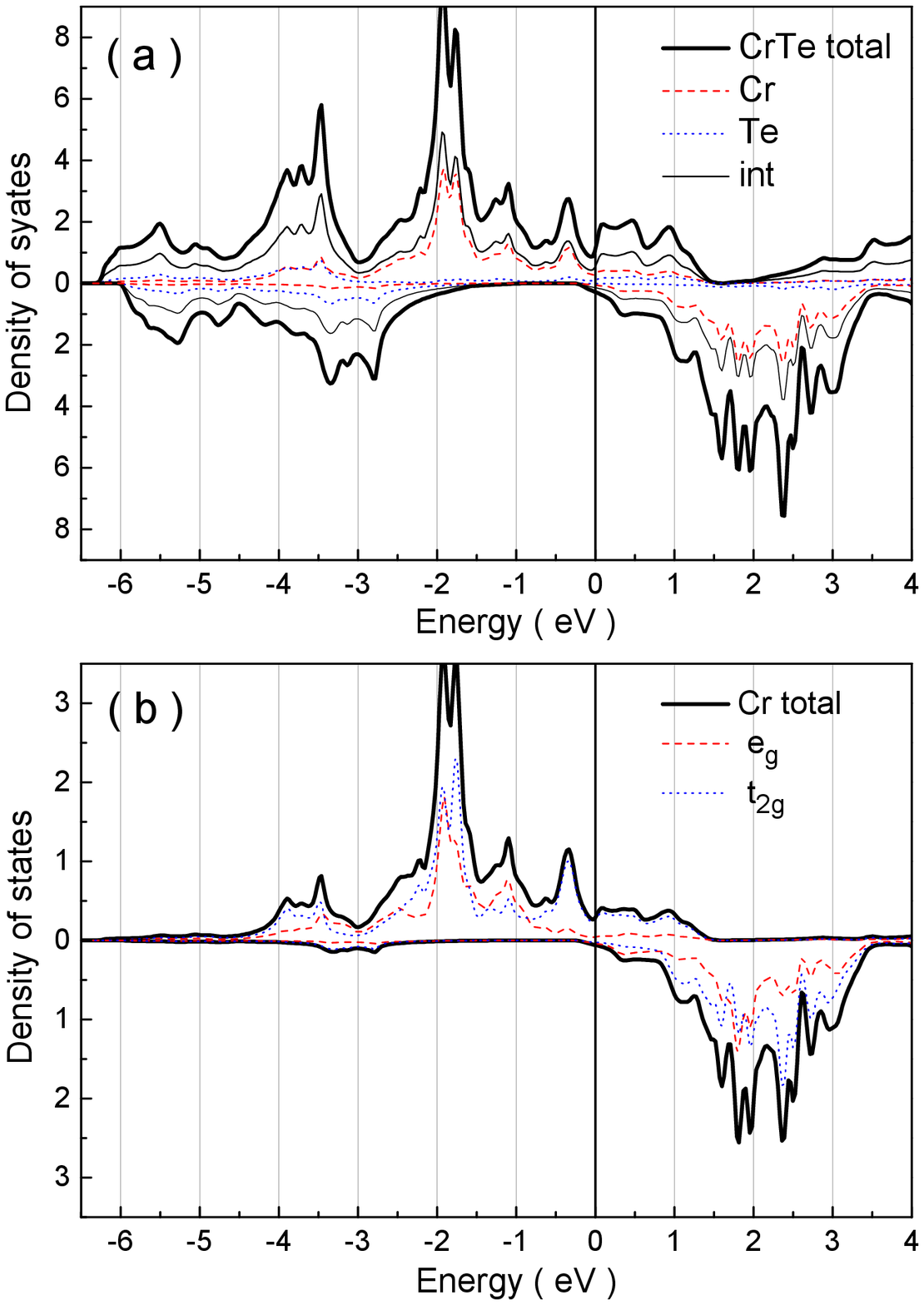}
\caption{Spin-polarized total and partial density of states of the nickel-arsenide CrTe in the ferromagnetic ground-state phase at the equilibrium lattice constants, calculated with mBJ functionals.}\label{edge}
\end{figure}

\subsection{Strains and electronic phase transition}

To explore the potential half-metallicity of nickel-arsenide ferromagnetic CrTe phase, we apply an in-plane biaxial compressive stress on it. As a result, this compressive stress will produce a decrease of the in-plane lattice constant $a$ and an increase of the out-of-plane $c$. These imply that the in-plane strain is compressive and the out-of-plane strain tensile. We change the in-plane strain so that the in-plane lattice constant $a$ is expanded from 1\% to 7\% with respect to the equilibrium constant $a_0$. The optimized $c$ value is obtained by minimizing the total energy as a function of $c$, which implies that the z-axis stress is made zero. With this series of strained lattice constants $a$ and $c$, we can calculate the bond angles, total magnetic moment, and total energy as functions of $a$ or $\Delta a/a_0$. Here, the total moment and total energy are per formula unit. To confirm the half-metallic property \cite{hm,hm-lbg}, we should investigate the half-metallic gap for the strained CrTe. Here, the half-metallic gap is defined as the smaller one of absolute energy differences between the Fermi energy and the conduction band bottom or the valence band top in the minority spin \cite{hm,hm-lbg}. All the calculated results are summarized in Table I. The calculated total magnetic moment is less than 4.0000$\mu_B$ when $\Delta a/a_0$ is less than 4.8\%, but when $\Delta a/a_0$ reaches to 4.8\%, the total magnetic moment becomes $4\mu_B$, an integer in magneton, which is a symbol of half-metallic feature \cite{hm,hm-lbg}. Additionally, the half-metallic gap becomes nonzero when $\Delta a/a_0$ is larger than 4.8\%. Also presented in Table I are the magnetic energies, $\Delta E_m$, for all the $\Delta a/a_0$ values. $\Delta E_m$, defined as the relative total energy  per formula unit of the antiferromagentic order, remains approximately 0.21 eV when $\Delta a/a_0$ increases to 7\%. This means that the ferromagnetic phase is stable against antiferromagnetic fluctuations and the Curie temperature should not be affected by the strain effects.

\begin{table}[!h]
\caption{The strain dependence of the out-of-plane lattice constant ($c$ in \AA), the two bond angles ($\alpha_1$ and $\alpha_2$ in $^\circ$), the magnetic moment ($M$ in $\mu_B$), the relative total energy ($\Delta E$ in eV), the half-metallic gap ($G_{hm}$ in eV), and the magnetic energy ($\Delta E_m$ in eV).}
\begin{ruledtabular}
\begin{tabular}{cccccccc}
$\Delta a/a_0$ & $c$	  &  $\alpha_1$  & $\alpha_2$	& $M$ & $\Delta E$ & $G_{hm}$  & $\Delta E_m$ \\ \hline
0\%	           & 6.292  & 92.6	& 66.8	& 3.966 & 0     &   -     &  0.214 \\
1.0\%          & 6.229  & 93.3  & 65.8  & 3.975 & 0.005 &   -     &  0.211 \\
2.0\%          & 6.167  & 94.0  & 64.8  & 3.984 & 0.014 &   -     &  0.207 \\
3.0\%          & 6.118  & 94.6  & 63.9  & 3.995 & 0.028 &   -     &  0.208 \\
4.0\%          & 6.071	& 95.2	& 63.0	& 3.999 & 0.045 &   -     &  0.207 \\
4.5\%          & 6.060	& 95.5	& 62.6	& 4.000 & 0.056 &   -     &  0.208 \\
4.8\%          & 6.057  & 95.6  & 62.4  & 4.000 & 0.064 &   0.0   &  0.207 \\
5.0\%          & 6.054  & 95.7  & 62.3  & 4.000 & 0.070 &   0.20  &  0.207 \\
5.5\%          & 6.048  & 95.9  & 62.0  & 4.000 & 0.083 &   0.30  &  0.209 \\
6.0\%          & 6.043  & 96.0  & 61.7  & 4.000 & 0.097 &   0.36  &  0.210 \\
7.0\%          & 6.032  & 96.4  & 61.2  & 4.000 & 0.134 &   0.50  &  0.207
\end{tabular}
\end{ruledtabular}
\end{table}

\begin{figure}[!htbp]
\centering  % Requires \usepackage{graphicx}
\includegraphics[clip, width=7cm]{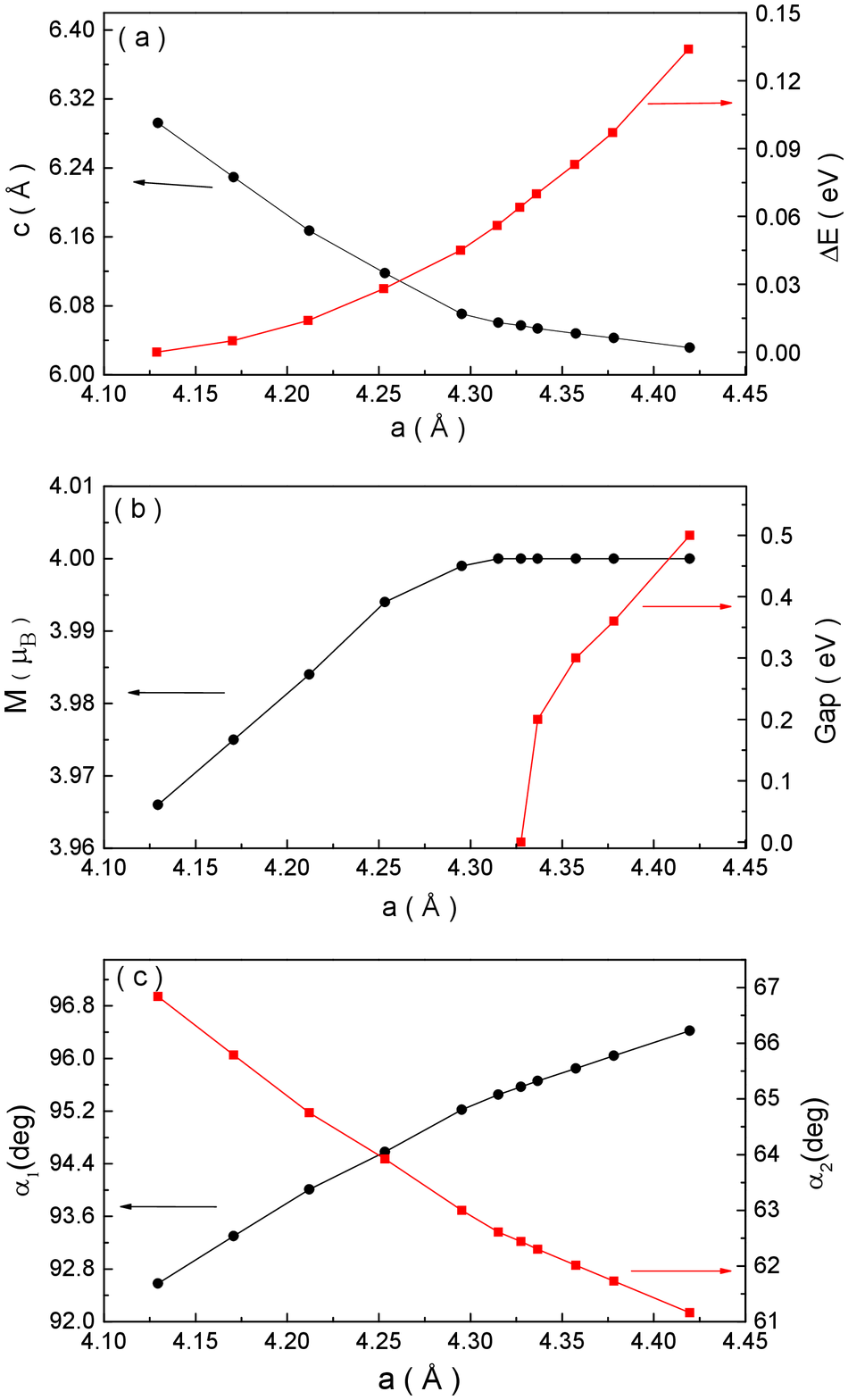}
\caption{Strain dependence of lattice constant $c$ and the total energy $\Delta E$ (a), magnetic moment and half-metallic gap (b), and bond angles (c).}\label{edge}
\end{figure}

We present in Fig. 2 the strain dependences of the z-axis lattice constant $c$, relative total energy $\Delta E$, magnteic moment $M$, half-metallic gap, and bond angles ($\alpha_1$ and $\alpha_2$) of the strained hexagonal CrTe. The $c$ decreases with $a$ increasing, which is a usual behavior, hindering the change of the volume. The relative total energy $\Delta E$ can be described with a parabola, reflecting that such strains are still in the linear regime. It means that a half-metallic phase is achieved at $\Delta a/a_0=4.8$\% because the magnetic moment reaches 4$\mu_B$ and the nonzero half-metallic gap appears at the strain \cite{hm,hm-lbg}. This half-metallic property, realized in the hexagonal ground-state phase, is different from that in meta-stable cubic CrTe\cite{crte-lbg,aaa}. This biaxial stress approach with $\Delta a/a_0=4.8$\% can be experimentally realized  by growing the hexagonal CrTe thin films on appropriate tensile substrates, being in contrast to isotropic expansion of about 10\% which should be unrealistic\cite{aaa}. The half-metallic ferromagnetism implies that the full spin polarization can be made. $\alpha_1$ increases with $a$ increasing, but $\alpha_2$ decreases. These trends are essentially caused by the decreasing of $c$, and thus are indirectly caused by the increasing of $a$, because the increasing of $a$ alone does not change neither $\alpha_1$ nor $\alpha_2$.

\subsection{Half-metallic phase and full spin polarization}

\begin{figure}[!htbp]
\centering  % Requires \usepackage{graphicx}
\includegraphics[clip, width=7cm]{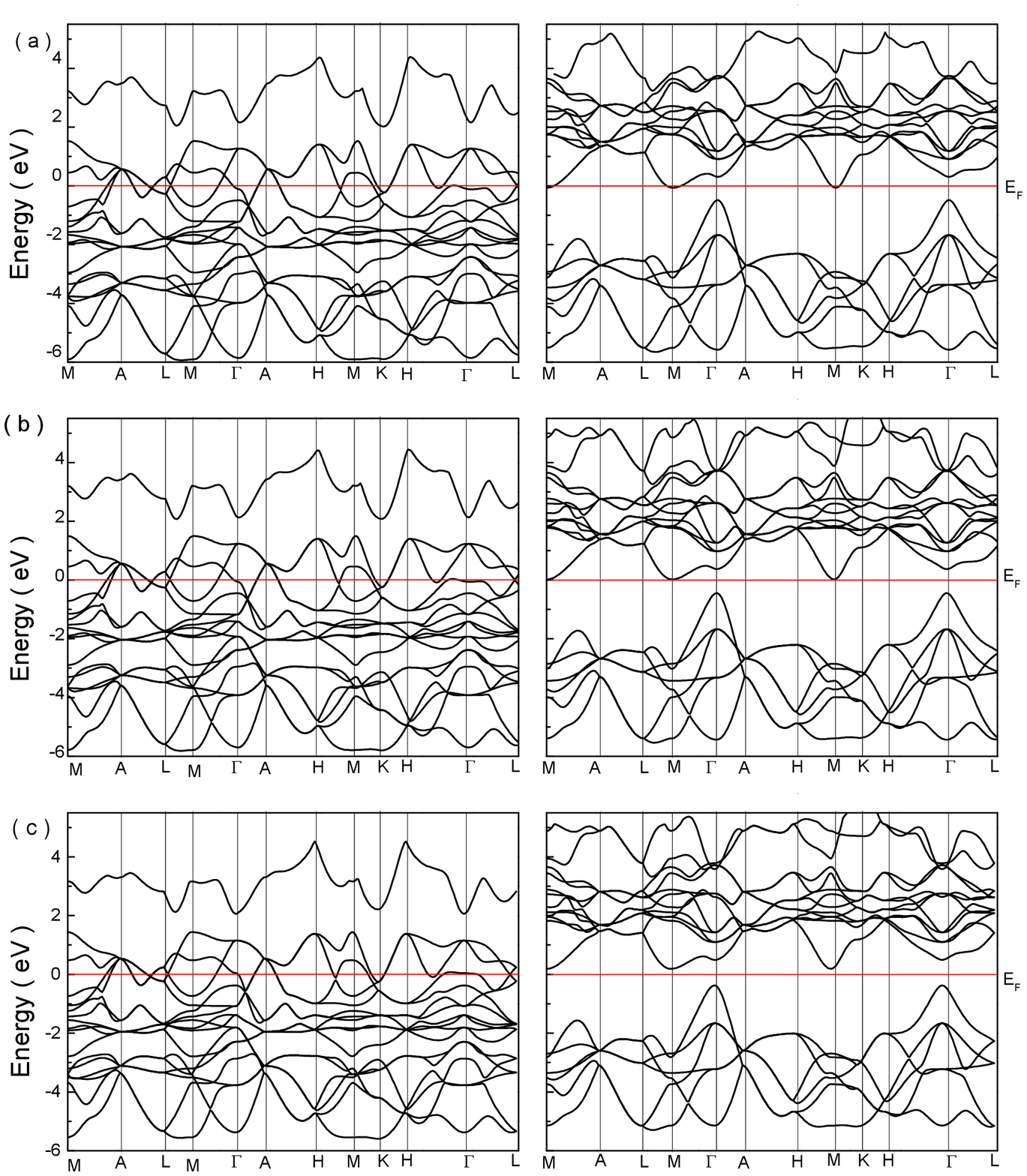}
\caption{Spin-polarized band structures of the hexagonal ferromagnetic CrTe phase for $\Delta a/a_0$ = 4.0\% (a), 4.8\% (b), and 7.0\% (c) calculated with mBJ.}\label{edge}
\end{figure}

\begin{figure}[!tbp]
\centering  % Requires \usepackage{graphicx}
\includegraphics[clip, width=7cm]{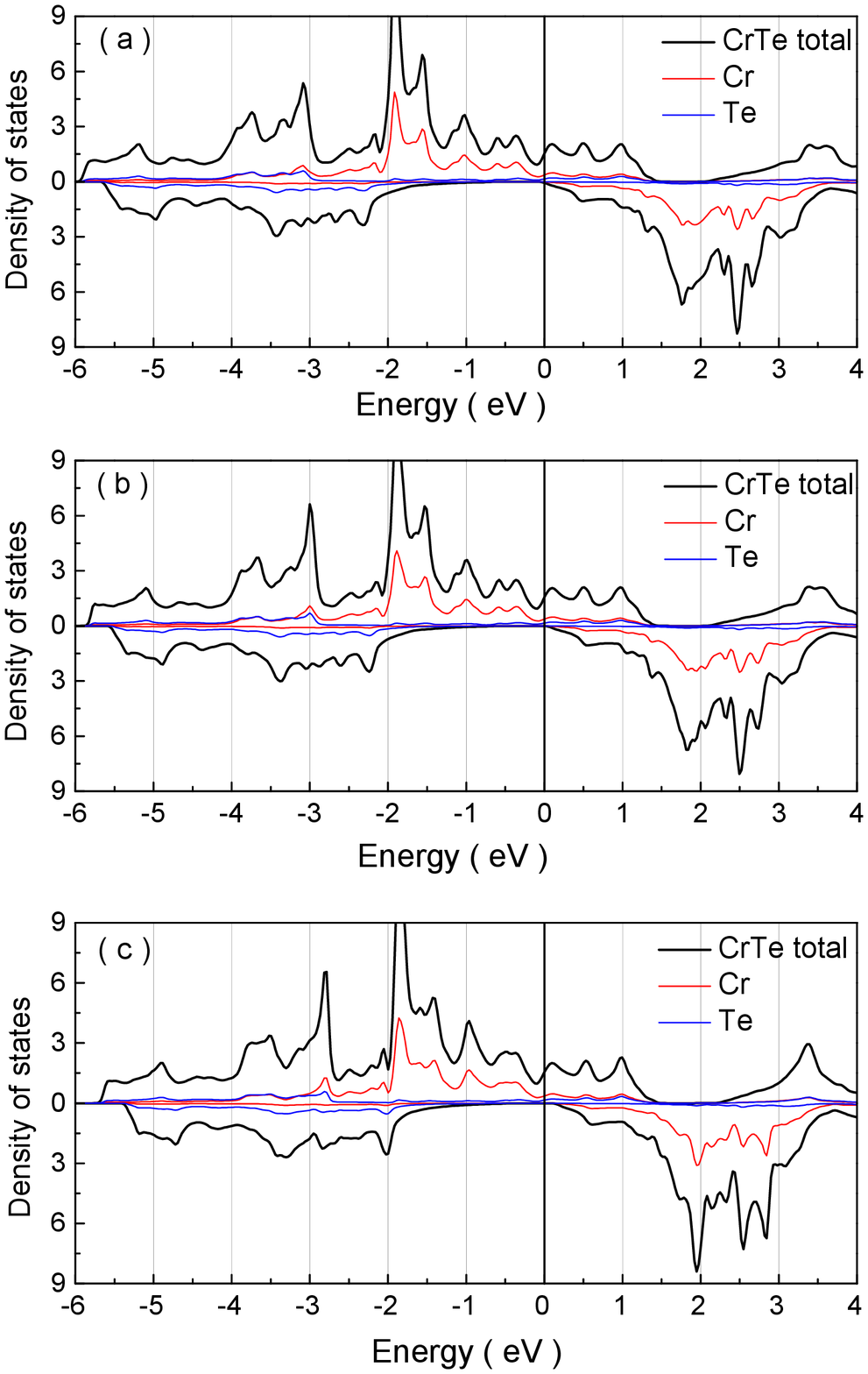}
\caption{Spin-polarized total and partial density of states projected in the atomic spheres of the hexagonal ferromagnetic CrTe phase for $\Delta a/a_0$ = 4.0\% (a), 4.8\% (b), and 7.0\% (c) calculated with mBJ.}\label{edge}
\end{figure}

In order to show the changes of the electronic structure of the hexagonal CrTe over the conventional-to-half-metallic transition, we present the spin-resolved band structure and density of states for $\Delta a/a_0$ = 4.0\%, 4.8\%, and 7.0\% in Figs. 3 and 4, respectively. At the equilibrium lattice constants, the hexagonal CrTe is a conventional ferromagnetic metal with four Cr d electrons, with the Te p orbitals being fully filled, and there is a small gap between -1.2 and -0.3 eV in minority-spin channel. The conduction band bottom in minority-spin channel, mainly from Cr d states, is contributed by the Brillouin zone part near the M point. At $\Delta a/a_0$ = 4.0\%, the hexagonal CrTe is still in conventional ferromagnetic phase, but it transits to a half-metallic ferromagnetic phase when $\Delta a/a_0$ is larger than 4.8\%, which can be clearly seen in both of the band structures and the density of states. The minority-spin gap actually becomes a little smaller, but the Fermi level gets in the gap at $\Delta a/a_0$ = 4.8\%. This is caused by the in-plane contraction of the bands, especially in the $\Gamma$-M direction, due to the in-plane expansion of the hexagonal CrTe.

\begin{figure}[!tbp]
\centering  % Requires \usepackage{graphicx}
\includegraphics[clip, width=7cm]{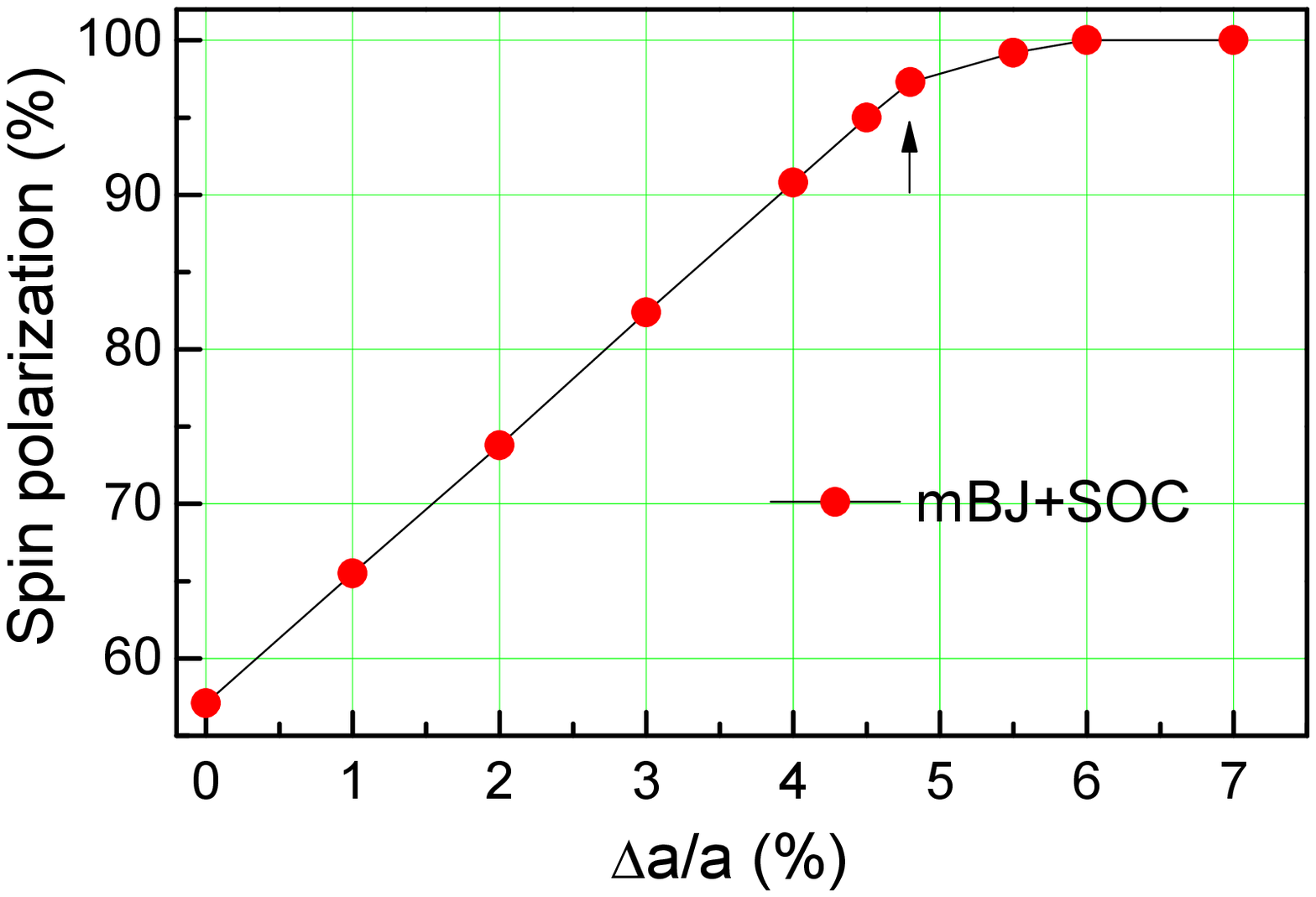}
\caption{Strain-dependent spin polarization at the Fermi level calculated with both mBJ and SOC. The arrow denotes the transition point without considering the spin-orbit coupling. }\label{edge}
\end{figure}

It is also important to investigate the effect of the spin-orbit coupling on the spin polarization at the Fermi level because it usually decreases spin polarization. For the $a$ strain up to 7\%, we calculate the band structures with mBJ functional and the spin-orbit coupling taken into account, and present in Fig. 5 the spin polarization values at the Fermi level. It is clear that the spin polarization almost linearly increases with $\Delta a/a_0$ up to 4.5\%, and reaches to 100.0\% from $\Delta a/a_0$ = 6.0\% on. The spin-orbit coupling substantially reduces the spin polarization at $\Delta a/a_0$ = 4.5$\sim$6.0\%, and thus postpones the transition point to $\Delta a/a_0$ = 6.0\%.

The biaxial tensile stress actually produces the hybrid strain that is tensile in the xy plane and compressive in the z axis. This stress can be realized by epitaxially growing the hexagonal CrTe on some suitable semiconductor substrates with hexagonal surfaces and larger surface lattice constants. The hybrid strain can cause the necessary in-plane contraction of the bands and thereby enhance the spin polarization at the Fermi level or even make the full spin polarization. This mechanism is different from those of half-metallic properties in oxides\cite{cro2q,lsmo,mfro}, imter-metallic compound\cite{mrg}, and transition-metal-doped graphene system\cite{v-graphene}.

\section{Conclusion}

In summary, we have made the hexagonal NiAs structure (the ground-state phase) of CrTe become half-metallic at an in-plane strain of 4.8\% by applying realizable biaxial tensile stress on it. Reliable electronic structure has been achieved by using the improved exchange functional (mBJ) and taking the spin-orbit coupling into account. With the in-plane strain increasing, the spin polarization is 97\% at the transition point and then reaches 100.0\% when the strain is at 6.0\%. The mechanism of the half-metallic phase is that the tensile in-plane strain makes the energy bands contract in the $\Gamma-M$ direction in the Brillouin zone and thereby shifts the Fermi level in the gap in minority-spin channel. Because it should not be difficult to realize this hexagonal ground-state phase and the tensile biaxial stress, these results should be useful to stipulate experimental effort to fabricate high-quality epitaxial thin films on suitable substrates for spintronics applications.

\begin{acknowledgments}
This work is supported by the Nature Science Foundation of China (Grant No. 11174359 and No. 11574366), by the Department of Science and Technology of China (Grant No. 2016YFA0300701), and by the Strategic Priority Research Program of the Chinese Academy of Sciences (Grant No.XDB07000000).
\end{acknowledgments}

\end{document}